\documentclass[aps,prc,preprint,showpacs,superscriptaddress]{revtex4}

\usepackage[dvips]{graphicx,color}
\usepackage{dcolumn}
\usepackage{bm}

\def\pmb#1{\setbox0=\hbox{#1}%
  \kern-.025em\copy0\kern-\wd0 
  \kern.05em\copy0\kern-\wd0
  \kern-.025em\raise.0433em\box0 }

\newcommand{\Figuretable}[1]{%
  \begin{center} --------- {\bf #1} --------- \\ \end{center}} 
\def\lambdabar{\protect\@lambdabar}
\def\@lambdabar{%
\relax
\bgroup
\def\@tempa{\hbox{\raise.73\ht0
\hbox to0pt{\kern.25\wd0\vrule width.5\wd0
height.1pt depth.1pt\hss}\box0}}%
\mathchoice{\setbox0\hbox{$\displaystyle\lambda$}\@tempa}%
{\setbox0\hbox{$\textstyle\lambda$}\@tempa}%
{\setbox0\hbox{$\scriptstyle\lambda$}\@tempa}%
{\setbox0\hbox{$\scriptscriptstyle\lambda$}\@tempa}%
\egroup
}

\begin{document}

\preprint{}

\title{
Feasibility of extracting a $\bm\Sigma^-$ admixture probability \\
in the neutron-rich $^{10}_{~\Lambda}$Li hypernucleus
}

\author{T.~Harada\footnote
{FAX: +81-72-825-4689; Telephone:+81-72-824-1131 (ext.4584);
E-mail:harada@isc.osakac.ac.jp}
}%


\author{A.~Umeya}%

\affiliation{%
Research Center for Physics and Mathematics,
Osaka Electro-Communication University, Neyagawa, Osaka, 572-8530, Japan
}

\author{Y.~Hirabayashi}%

\affiliation{%
Information Initiative Center, 
Hokkaido University, Sapporo, 060-0811, Japan
}

\date{\today}

\begin{abstract}

We theoretically examine production of the neutron-rich 
$^{10}_{~\Lambda}$Li hypernucleus by a double-charge exchange 
($\pi^-$,~$K^+$) reaction on a $^{10}$B target 
with distorted-wave impulse approximation calculations. 
We calculate the inclusive spectrum at the incident momentum 1.20 GeV/c 
by a one-step mechanism $\pi^-p \to K^+ \Sigma^-$ via $\Sigma^-$ doorways 
caused by a $\Sigma^-p \leftrightarrow \Lambda n$ coupling.
The resultant spectrum can explain the magnitude of the recent 
experimental data, 
so that the $\Sigma^-$ admixture probability in $^{10}_{~\Lambda}$Li
is found to be the order of 10$^{-1}$ \%. 
The ($\pi^-$,~$K^+$) reaction provides a capability 
of extracting properties of wave functions with 
$\Lambda$-$\Sigma$ coupling effects in neutron-rich nuclei, 
as well as the reaction mechanism. 

\end{abstract}
\maketitle


It has been discussed that a study of strangeness in nuclei 
would provide new information on nuclear physics 
and astrophysics \cite{NPA804}. 
The presence of hyperons in high-density nuclear medium significantly 
affects the maximal mass of neutron stars and compact stars, 
because it makes 
the Equation of State (EoS) soften \cite{Balberg97}.  
The negatively charged $\Sigma^-$ hyperon 
was expected to play an essential role in description 
of neutron stars, 
whereas the baryon fraction is found to depend on properties of 
hypernuclear potentials in neutron stars.  
On the other hand, several theoretical studies \cite{Friedman07} 
suggested that a repulsive component in $\Sigma^-$-nucleus 
potentials is needed to reproduce the observed spectra of 
($\pi^-$,~$K^+$) reactions on nuclear targets \cite{Saha04}
and also the strong level-shifts and widths of the $\Sigma^-$ 
atomic $X$-ray data. 
This repulsion originates from the $\Sigma N$ $T=3/2$, $^3S_1$ 
channel which corresponds to a quark Pauli-forbidden 
state in the baryon-baryon system \cite{Fujiwara07}.
However, since a strong $\Sigma^- p \to \Lambda n$ conversion 
occurs at a nuclear surface, 
it is difficult to extract the nature of the $\Sigma^-$ hyperon 
in nuclear medium from such experimental data on 
nuclear targets. 

One of the most promising subjects to examine 
the hypernuclear potential in a neutron-excess environment 
is a study of neutron-rich $\Lambda$ hypernuclei \cite{Majling95}. 
The $\Lambda$ hyperon in nuclei is known to act as a nuclear ``glue'', 
and can often make the system bound 
even if a core-nucleus is unbound, e.g., $^6_\Lambda$He.
In addition, it is suggested that in $s$-shell $\Lambda$ hypernuclei 
an attractive mechanism appears due to the $\Lambda$-$\Sigma$ 
coupling which is related to a three-body $\Lambda NN$ 
force \cite{Akaishi00}, 
and their $\Sigma$-mixing probabilities are 1-2 \%, 
as discussed in few-body calculations \cite{Nemura04}. 
This situation is found to be more coherently enhanced 
in the neutron-excess environment \cite{Myint00}.
Therefore, 
we believe that there are a lot of exotic neutron-rich 
$\Lambda$ hypernuclei beyond the neutron-drip line. 

\Figuretable{FIG. 1}

\begin{figure}[ht]
  \begin{center}
  \includegraphics[width=0.6\linewidth]{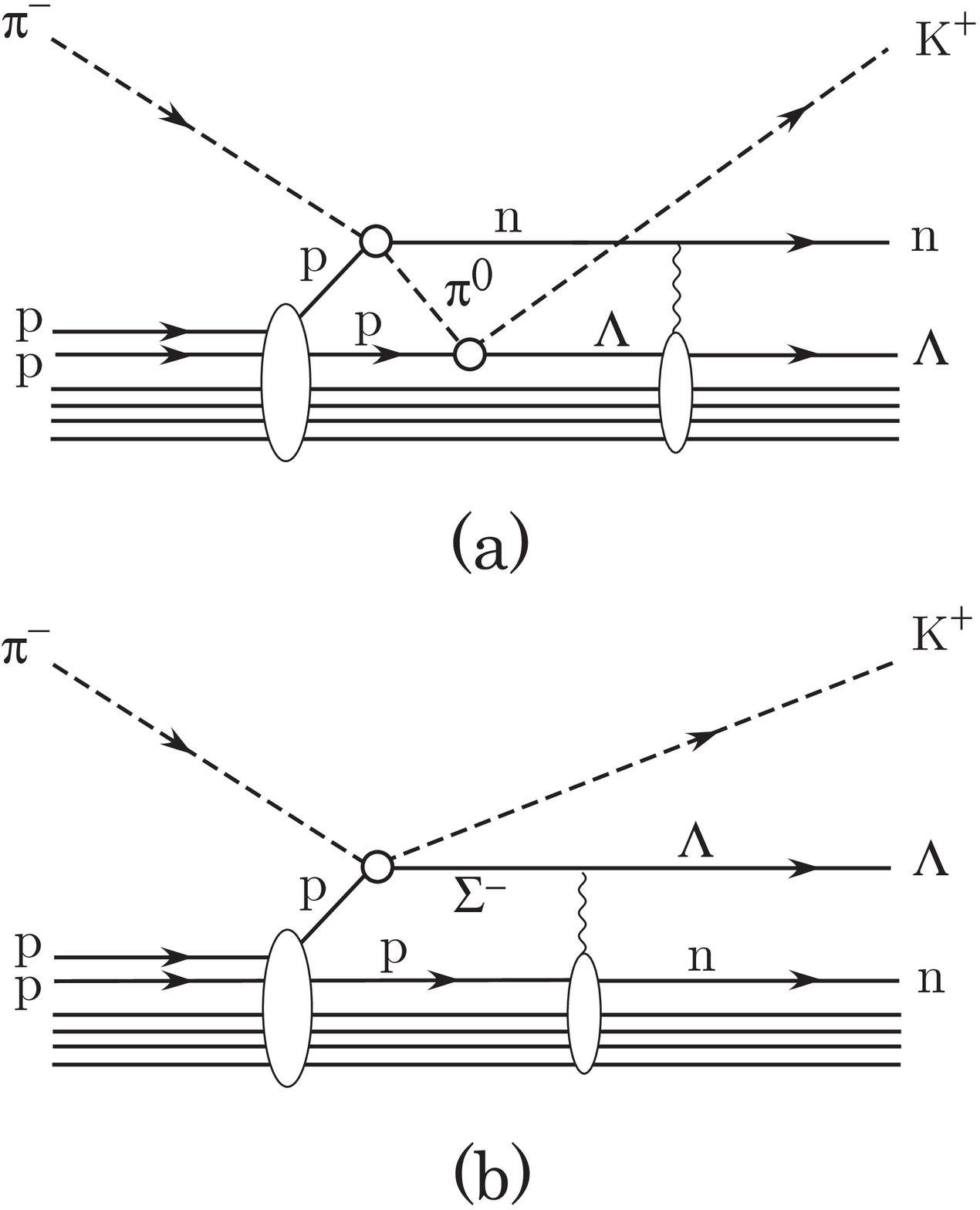}
  \caption{\label{fig:1}
  Diagrams for DCX nuclear ($\pi^-$,~$K^+$) reactions, 
  leading to production of $\Lambda$ hypernuclear states:  
  (a) a two-step mechanism, $\pi^-p$ $\to$ $\pi^0n$ followed by 
  $\pi^0p$ $\to$ $K^+\Lambda$, and (b) a one-step mechanism, 
  $\pi^-p$ $\to$ $K^+ \Sigma^-$ via $\Sigma^-$ doorways caused by 
  the $\Sigma^-p$ $\leftrightarrow$ $\Lambda n$ coupling.
  }
  \end{center}
\end{figure}

The experimental attempts to produce neutron-rich $\Lambda$ hypernuclei
were carried out by reactions based on a double-charge exchange (DCX) 
mechanism, as ($K^-_{\rm Stopped}$,~$\pi^+$) \cite{Kubota96,Agnello06} 
and ($\pi^-$,~$K^+$) \cite{Saha05}.
Further experiments in the nuclear ($\pi^-$,~$K^+$) reactions are also 
planned at J-PARC \cite{Sakaguchi07}.
The production of the neutron-rich $\Lambda$ hypernuclei by 
the DCX reaction ($\pi^-$,~$K^+$) would 
conventionally proceed by a two-step mechanism of the meson charge-exchange, 
$\pi^-p$ $\to$ $\pi^0n$ followed by $\pi^0p$ $\to$ $K^+\Lambda$, 
as shown in Fig.~\ref{fig:1}(a), or $\pi^-p$ $\to$ $K^0\Lambda$ 
followed by $K^0p$ $\to$ $K^+n$. 
Another exotic mechanism is a one-step process, 
$\pi^-p$ $\to$ $K^+ \Sigma^-$ via $\Sigma^-$ doorways caused by 
the $\Sigma^-p$ $\leftrightarrow$ $\Lambda n$ coupling
in $\Lambda$ hypernuclei, as shown in Fig.~\ref{fig:1}(b).  
Tretyakova and Lanskoy \cite{Tretyakova03} theoretically found
that the two-step mechanism in the $^{10}$B($\pi^-$,~$K^+$) 
reaction is more dominant compared to the one-step one, 
where the $\Sigma^-$ admixture probability is 
as small as the order of 10$^{-2}$ \%.
Thus they claimed that 
the magnitude of the cross section of the $^{10}_{~\Lambda}$Li 
bound state in the ($\pi^-$,~$K^+$) 
reaction is as large as 38-67 nb/sr at 
the incident momentum $p_\pi=$ 1.05 GeV/c (0$^\circ$), 
where the cross section of the conventional ($\pi^+$, $K^+$) reaction 
on nuclear targets is at its maximum \cite{Bando90}. 

Recently, Saha et al. \cite{Saha05} have performed the 
first measurement of a significant yield 
for the $^{10}_{~\Lambda}$Li hypernucleus in 
($\pi^-$,~$K^+$) reactions on a $^{10}$B target, 
whereas no clear peak has been observed with 
the lack of the experimental statistics. 
The data show that the absolute cross section
for $^{10}_{~\Lambda}$Li at 1.20 GeV/c ($d\sigma/d\Omega \sim$ 11 nb/sr) 
is twice larger than that at 1.05 GeV/c ($d\sigma/d\Omega \sim$ 6 nb/sr). 
This incident-momentum dependence of $d\sigma/d\Omega$ exhibits
a trend in the opposite direction for the theoretical prediction 
of Ref.~\cite{Tretyakova03}.
This might mean that the one-step mechanism is favored rather 
than the two-step mechanism, as pointed out in Ref.~\cite{Saha05}.

In this paper, we theoretically investigate production of 
the neutron-rich $^{10}_{~\Lambda}$Li hypernucleus by 
the DCX ($\pi^-$,~$K^+$) reaction on a $^{10}$B target 
at 1.20 GeV/c,  
within a distorted-wave impulse approximation (DWIA). 
In order to understand the mechanism of this reaction, 
we focus on the $\Lambda$ spectrum populated by the one-step mechanism, 
$\pi^- p \to K^+ \Sigma^-$ via $\Sigma^-$ doorways 
due to the $\Sigma^- p$ $\leftrightarrow$ $\Lambda n$ coupling.
This is the first attempt to extract the probability of 
the $\Sigma^-$ admixture in a neutron-rich $\Lambda$ hypernucleus 
from available experimental data phenomenologically. 
We also discuss a small contribution of the two-step
processes in the ($\pi^-$,~$K^+$) reactions within 
the eikonal approximation.

Now let us consider the DCX ($\pi^-$,~$K^+$) reaction on 
the $^{10}$B target within the DWIA.
In order to fully describe the one-step process via $\Sigma^-$ 
doorways, as shown in Fig.~\ref{fig:1}(b), 
we perform a $\Lambda$-$\Sigma$ coupled-channel 
calculation \cite{Harada98}, 
evaluating the production cross section of $\Lambda$ hypernuclear 
states in $^{10}_{~\Lambda}{\rm Li}$. 
We assume a two-channel coupled wave function for simplicity,  
which is given by 
\begin{equation}
| ^{10}_\Lambda{\rm Li} \rangle
= \varphi_\Lambda({\bm r}) \, | ^{9}{\rm Li}\otimes \Lambda \rangle 
+  \varphi_\Sigma({\bm r}) \, | ^{9}{\rm Be}^* \otimes \Sigma^- \rangle, 
\label{eqn:e1}
\end{equation}
where $\langle {\varphi}_\Lambda| \varphi_\Lambda \rangle
+\langle {\varphi}_\Sigma| \varphi_\Sigma \rangle=1$, and 
${\bm r}$ denotes a relative coordinate between the core-nucleus 
and the hyperon. 
The probability of the $\Sigma^-$ admixture in the $\Lambda$ 
hypernucleus can be obtained by 
$P_{\Sigma^-}=\langle {\varphi}_\Sigma| \varphi_\Sigma \rangle$.
It should be noticed that 
the core-excited state ($^9$Be$^{*}$) in the $\Sigma^-$ channel 
is assumed to be a one effective state which can be fully 
coupled with the $^{9}$Li core state via the $\Lambda$-$\Sigma$ 
coupling, 
rather than the $^9$Be(${3 \over 2}^-$;${1 \over 2}$) ground state. 
Thus, we assume $\Delta M =$ 80 MeV effectively for 
the threshold-energy difference between $^{9}$Li+$\Lambda$ 
and $^{9}$Be$^*$+$\Sigma^-$ channels.
The mixed $\Sigma^-$-hyperon in $^{10}_{~\Lambda}$Li 
is regarded as a deeply bound particle
in the nucleus, where the $\pi^-p \to K^+ \Sigma^-$ transition 
takes place under an energy-off-shell condition. 

In order to calculate the nuclear ($\pi^-$,~$K^+$) spectrum, 
we employ the Green's function method \cite{Morimatsu94}, 
which is one of the most powerful treatments in a calculation 
of the spectrum which includes not only bound states but also continuum states with an 
absorptive potential for spreading components.
The complete Green's function ${\bm G}$ describes 
all information concerning 
(${^9}{\rm Li} \otimes \Lambda$)+(${^9}{\rm Be}^* \otimes \Sigma^-$) 
coupled-channel dynamics. 
We obtain it by solving the following equation
with the hyperon-nucleus potential ${\bm U}$ numerically:
\begin{equation}
{\bm G}={\bm G}^{(0)}
+{\bm G}^{(0)}{\bm U}{\bm G},
\label{eqn:e2}
\end{equation}
where 
\begin{equation}
{\bm G}=\left(
\begin{array}{cc}
    {G}_{\Lambda\Lambda}   & {G}_{\Lambda\Sigma}  \\
    {G}_{\Sigma\Lambda} & {G}_{\Sigma\Sigma}   
\end{array}
\right), \quad
{\bm U}=\left(
\begin{array}{cc}
    {U}_{\Lambda\Lambda}   & {U}_{\Lambda\Sigma}           \\
    {U}_{\Sigma\Lambda}         & {U}_{\Sigma\Sigma}   
\end{array}
\right), 
\label{eqn:e3}
\end{equation}
and ${\bm G}^{(0)}$ is a free Green's function. 
By the complete Green's function, 
the inclusive $K^+$ double-differential lab cross section 
of $\Lambda$ production on a nuclear target with a spin 
$J_i$ (its $z$-component $M_i$) \cite{Harada98} 
by the one-step mechanism, 
$\pi^- p \to K^+ \Sigma^-$ via $\Sigma^-$ 
doorways, is given by
\begin{equation}
{{d^2\sigma} 
\over {d\Omega_{K} dE_{K}} } 
 = 
\beta {1 \over {[J_i]}} \sum_{M_i}
(-){1 \over \pi}{\rm Im} \sum_{\alpha \alpha'} 
\left\langle
F_{\Sigma}^{\alpha \, \dagger} {G}_{\Sigma\Sigma}^{\alpha\alpha'}
F_{\Sigma}^{\alpha'} 
\right\rangle,
\label{eqn:e4}
\end{equation}
where a production amplitude 
\begin{equation}
  F_{\Sigma}^{\alpha} = {\overline{f}}_{\pi^-p \to K^+\Sigma^-}
  \chi_{{\bm p}_{K}}^{(-) \ast}
  \chi_{{\bm p}_{\pi}}^{(+)} 
  \langle \alpha \, | \hat{\mit\psi}_p | \, i \rangle,
\label{eqn:e5}
\end{equation}
$[J_i]=2J_i+1$, and the kinematical factor $\beta$ expresses 
the translation 
from the two-body $\pi^-$-$p$ lab system to the $\pi^-$-$^{10}$B 
lab system. 
$\overline{f}_{\pi^-p \to K^+\Sigma^-}$ is a  
Fermi-averaged amplitude for the $\pi^-p \to K^+\Sigma^-$ reaction 
in nuclear medium, and 
$\chi_{{\bm p}_{K}}^{(-)}$ and $\chi_{{\bm p}_{\pi}}^{(+)}$ 
are the distorted waves for outgoing $K^+$ and 
incoming $\pi^-$ mesons, respectively, 
taken into account the recoil effects \cite{Harada04}. 
$\langle \alpha \, | \hat{\mit\psi}_p  | \,  i \rangle$ 
is a hole-state wave function for a struck proton in the target, 
where $\alpha$ denotes the complete set of eigenstates for the system.
The inclusive $\Lambda$ spectrum in Eq.(\ref{eqn:e4}) can 
be decomposed into different physical 
processes \cite{Morimatsu94,Harada98}, by using the identity
\begin{eqnarray}
{\rm Im}(F_{\Sigma}^\dagger {G}_{\Sigma\Sigma}F_{\Sigma})
&=& F_{\Sigma}^\dagger{\Omega}^{(-){\dagger}}
({\rm Im}{G}_\Lambda^{(0)}){\Omega}^{(-)}F_{\Sigma} \nonumber \\
&+& F_{\Sigma}^\dagger{G}_{\Sigma\Lambda}^{\dagger}
({\rm Im}{U}_\Lambda){G}_{\Lambda\Sigma} F_{\Sigma} \nonumber \\
&+& F_{\Sigma}^\dagger{G}_{\Sigma\Sigma}^{\dagger}
({\rm Im}{U}_\Sigma){G}_{\Sigma\Sigma} F_{\Sigma},
\label{eqn:e6}
\end{eqnarray}
where ${\Omega}^{(-)}$ is the M{\"o}ller wave operator. 

The diagonal (optical) potentials for ${\bm U}$ in Eq.~(\ref{eqn:e3})
are given by the Woods-Saxon (WS) form: 
\begin{equation}
U_{Y}(r)= (V_{Y}+iW_{Y}g(E_\Lambda))f(r)
\label{eqn:e7}
\end{equation} 
for $Y=$ $\Lambda$ or $\Sigma^-$, 
where $f(r)=[1 + \exp{((r-R)/a)}]^{-1}$ 
with $a=$ 0.6 fm, $r_0=$ $1.088+0.395 A^{-2/3}$ fm 
and $R=r_0(A-1)^{1/3}=$ 2.42 fm for the mass number $A=10$ \cite{Millener88}. 
Here we used $V_\Lambda=$ $-$30 MeV for the 
$^9{\rm Li}\otimes\Lambda$ channel, and assumed 
$V_\Sigma=$ 0 MeV to describe the effective $\Sigma$ state of 
the $^9{\rm Be}^* \otimes \Sigma^-$ channel. 
The spreading imaginary potential, ${\rm Im}U_Y$, can describe 
complicated excited-states for ${^{10}_{~\Lambda}}{\rm Li}$; 
$g(E_\Lambda)$ is an energy-dependent function 
which linearly increases
from 0 at $E_\Lambda=$ 0 MeV to 1 at $E_\Lambda=$ 60 MeV, 
as often used in nuclear optical models.
The strength parameter $W_Y$ should be 
adjusted appropriately to reproduce the data.  
The coupling $\Lambda$-$\Sigma$ potential $U_{\Sigma\Lambda}$ 
in off-diagonal parts for ${\bm U}$ 
is written by
\begin{eqnarray}
U_{\Sigma\Lambda}(r) 
&=& \langle ^{9}{\rm Be}^*\otimes \Sigma^-| 
{1 \over \sqrt{3}}\sum_j v_{\Sigma\Lambda}({\bm r}_j,{\bm r}) 
{{\bm \tau}_j \cdot {\bm \phi}} \nonumber\\
&& \times 
| {^{9}{\rm Li}} \otimes \Lambda \rangle,
\label{eqn:e8}
\end{eqnarray}
where $v_{\Sigma\Lambda}({\bm r}_j,{\bm r})$ is a two-body 
$\Lambda N$-$\Sigma N$ potential including the spin-spin 
interaction, 
and ${\bm \tau}_j$ denotes the $j$-th nucleon 
isospin operator and ${\bm \phi}$ is defined as
$|\Sigma \rangle = {\bm \phi}|\Lambda\rangle$ 
in isospin space \cite{Auerbach87}.
Here we assumed 
$U_{\Sigma\Lambda}(r)= V_{\Sigma\Lambda}f(r)$ 
in a real potential for simplicity,
where $V_{\Sigma\Lambda}$ is an effective strength 
parameter.  
We will attempt to determine the values
of $W_\Sigma$ and $V_{\Sigma\Lambda}$ phenomenologically
by fitting to a spectral shape of the experimental data.

For the $^{10}$B($3^+$;0) target nucleus, 
we use single-particle wave functions for a proton,
which are calculated by a WS potential \cite{Bohr69} 
with $V^N_{0}=$ $-$61.36 MeV fitting to 
the charge radius of 2.45 fm \cite{Vries87}. 
Due to large momentum transfer 
$q \simeq$ 270-370 MeV/c in the ($\pi^-$,~$K^+$) reaction, 
we simplify the computational procedure for the distorted waves, 
$\chi_{{\bm p}_{K}}^{(-)}$ and $\chi_{{\bm p}_{\pi}}^{(+)}$, 
with the help of the eikonal approximation. 
In order to reduce ambiguities in the distorted-waves, 
we adopt the same parameters used in calculations for the
$\Lambda$ and $\Sigma^-$ quasi-free spectra 
in nuclear ($\pi^\mp$,~$K^+$) reactions \cite{Harada04}.
Here we used total cross sections of $\sigma_\pi$= 32 mb 
for $\pi^- N$ scattering and $\sigma_K$= 12 mb for $K^+ N$ one, 
and $\alpha_\pi = \alpha_K =$ 0, as the distortion 
parameters \cite{Harada04}. 
The Fermi-averaged amplitude ${\overline{f}}_{\pi^-p \to K^+\Sigma^-}$
is obtained by the optimal Fermi-averaging for the 
$\pi^- p \to K^+ \Sigma^-$ t-matrix \cite{Harada04};
we used 20 $\mu$b/sr as the lab cross section of 
$d\sigma/d\Omega = |{\overline{f}}_{\pi^-p \to K^+\Sigma^-}|^2$.

\Figuretable{FIG. 2}

\begin{figure}[ht]
  \begin{center}
  \includegraphics[width=0.9\linewidth]{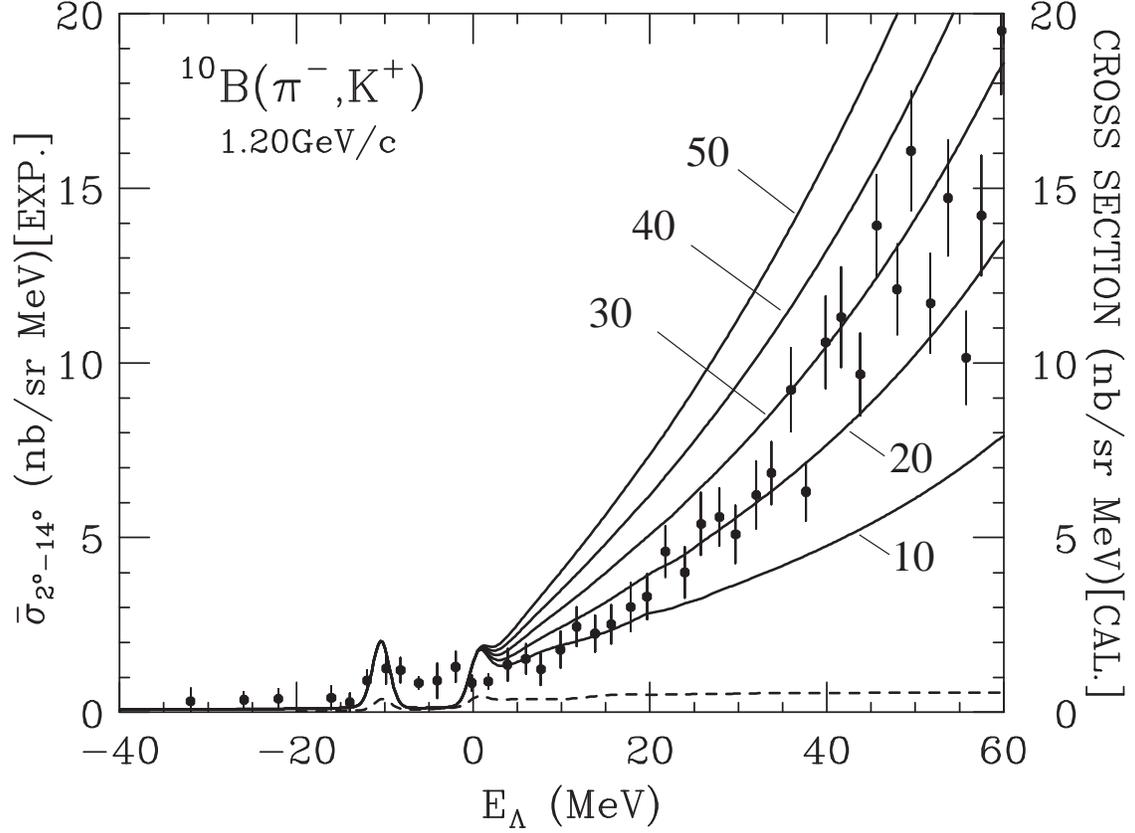}
  \caption{\label{fig:2}
  Calculated inclusive $\Lambda$ spectra by the one-step mechanism
  in the $^{10}$B($\pi^-$,~$K^+$) reaction at 
  $p_{\pi}=$ 1.20 GeV/c (6$^\circ$), 
  together with the experimental data \cite{Saha05}.
  The solid curves denote the $K^+$ spectra by $-W_{\Sigma}=$
  10, 20, 30, 40 and 50 MeV when $V_{\Sigma\Lambda}=$ 11 MeV ($P_\Sigma=$ 0.57 \%),
  with a detector resolution of 2.5 MeV FWHM. 
  The dashed curve denotes the inclusive $\Lambda$ spectrum by 
  the two-step mechanism. 
  }
  \end{center}
\end{figure}

Now let us examine the dependence of the spectral shape 
on two important parameters, $W_\Sigma$ and $V_{\Sigma\Lambda}$  
by comparing the calculated inclusive $\Lambda$ spectrum for
$^{10}_{~\Lambda}$Li with the data of $^{10}$B($\pi^-$,~$K^+$) 
experiments at KEK \cite{Saha04}.
The cross sections of the data are three orders of magnitude 
less than those for $^{10}_{~\Lambda}$B in $^{10}$B($\pi^+$,~$K^+$) reactions.
In Fig.~\ref{fig:2}, we show the calculated 
spectra by the one-step mechanism at $p_{\pi}=$ 1.20 GeV/c ($6^\circ$)
for the several values of $-W_{\Sigma}$  
when we use $V_{\Sigma\Lambda}=$ 11 MeV which leads to the 
$\Sigma^-$-mixing probability 
of $P_\Sigma=$ 0.57 \% in the $^{10}_{~\Lambda}{\rm Li}$ bound state.
We have the peak of the bound state with a 
[$0p_{3 \over 2}^{-1}s_{1 \over 2}^\Lambda$]$_{2^-}$ 
configuration at $E_\Lambda \simeq$ $-$10.0 MeV, 
and the peaks of the excited states with 
[$0p_{3 \over 2}^{-1}p_{{3 \over 2},{1 \over 2}}^\Lambda$]$_{3^+, 1^+}$ 
configurations at $E_\Lambda =$ 1-3 MeV.
Since the non-spin-flip processes with the large momentum transfer $q$
are known to dominate in the $\pi^-p \to K^+\Sigma^-$ reaction at 1.20 GeV/c, 
these spin-stretched states are mainly populated.
We find that the value of $-W_\Sigma$ significantly affects 
a shape of the $\Lambda$ spectrum for the continuum states 
which can be populated via $\Sigma^-p \to \Lambda n$ processes
in $^{9}$Be$^*$ together with the core-nucleus breakup; 
this $\Lambda$ strength mainly arises from a term of 
$G_{\Sigma\Sigma}^\dagger({\rm Im}U_\Sigma)G_{\Sigma\Sigma}$ 
in Eq.~(\ref{eqn:e6}).
We recognize that the calculated spectrum with 
$-W_\Sigma=$ 20-30 MeV can reproduce the shape of the data 
in the continuum region \cite{Saha05}, 
and these values of $-W_\Sigma$ are consistent with the analysis of 
$\Sigma^-$ production by the ($\pi^-$,~$K^+$) reactions 
\cite{Harada04}. 
Obviously, the parameter $W_\Sigma$ does not contribute to 
the spectrum of the bound state. 
It should be noticed that the contribution of the two-step processes
in the continuum spectrum is rather small, 
as the dashed curves shown in Fig.~\ref{fig:2}.

\Figuretable{FIG. 3}

\begin{figure}[ht]
  \begin{center}
 \includegraphics[width=0.9\linewidth]{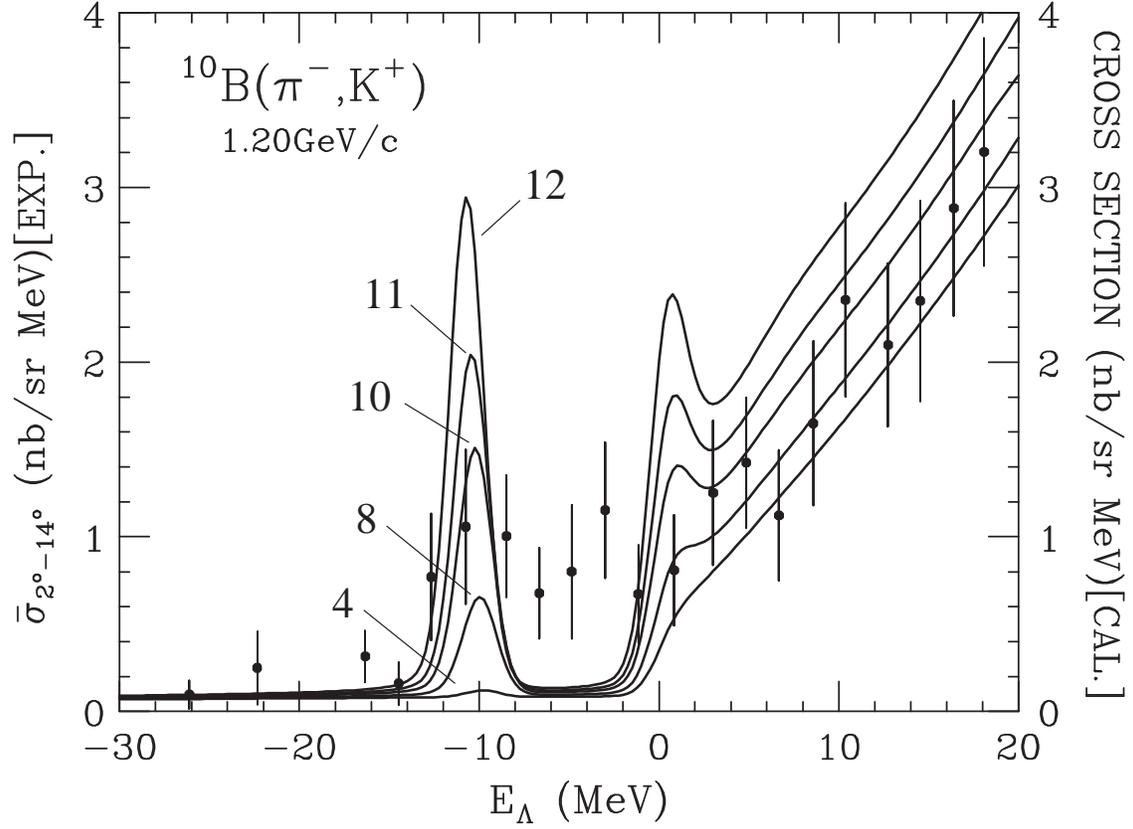}
  \caption{\label{fig:3}
  Calculated inclusive $\Lambda$ spectra by the one-step mechanism 
  near the $\Lambda$-threshold in the $^{10}$B($\pi^-$,~$K^+$) reaction 
  at 1.20 GeV/c (6$^\circ$), by changing $V_{\Sigma\Lambda}$ for the 
  $\Lambda$-$\Sigma$ coupling potential. 
  The experimental data are taken form Ref.~\cite{Saha05}. 
  The solid curves denote for $V_{\Sigma\Lambda}=$ $4$, $8$, $10$, $11$ 
  and $12$ MeV when $-W_{\Sigma}=$ 20 MeV,
  with a detector resolution of 2.5 MeV FWHM. 
  }
  \end{center}
\end{figure}

On the other hand, the $\Lambda$-$\Sigma$ coupling potential plays 
an essential role in the formation of the $\Lambda$ bound state. 
In Fig.~\ref{fig:3}, we show the dependence of the cross section for 
the bound state in $^{10}_{~\Lambda}{\rm Li}$ 
on the values of $V_{\Sigma\Lambda}$ when $-W_\Sigma=$ 20 MeV. 
We find that the calculated spectrum for the bound state 
is quite sensitive to $V_{\Sigma\Lambda}$;
when we use $V_{\Sigma\Lambda}=$ $4$, $8$, $10$, $11$ and $12$ MeV, 
the probabilities of the $\Sigma^-$ admixture in the $2^-$ bound state
are $P_{\Sigma^-}=$ $0.075$, $0.30$, $0.47$, $0.57$ and $0.68$ \%,
respectively. 
The positions of the peaks for $^{10}_{~\Lambda}{\rm Li}$ 
are slightly shifted downward by 
$\Delta E_\Lambda \simeq -\langle U_{\Sigma\Lambda} \rangle^2/\Delta M 
\simeq -\Delta M \cdot P_{\Sigma^-}$, e.g., 
$-$456 keV for $V_{\Sigma\Lambda}=$ 11 MeV, 
which is about 4-5 times larger than 
that of $^7_\Lambda$Li \cite{Millener08}. 
For the order of $V_{\Sigma\Lambda} =$ 10-12 MeV
($P_{\Sigma^-}=$ $0.47$-$0.68$ \%), the calculated spectra 
can fairly reproduce the data, 
whereas it is not appropriate to a detailed study of the structure 
of $^{10}_{~\Lambda}$Li 
because of the simple single-particle picture we adopted here.
Such a $\Sigma^-$ admixture seems to be consistent with 
the recent microscopic calculations \cite{Akaishi00,Nemura04,Umeya08}.
This consistency of $V_{\Sigma\Lambda}$ considerably enhances 
the reliability of our calculations. 
Consequently, the calculated spectrum by the one-step 
mechanism explains the $^{10}$B($\pi^-$,~$K^+$) data.
This fact implies that the one-step mechanism dominates 
in the ($\pi^-$,~$K^+$) reaction, 
and our calculations provide a capability of extracting a 
production mechanism from the data of this reaction. 
Some discrepancy between the results and the data 
in the bound-state region surrounding 
$E_\Lambda \simeq$ $-$5 MeV might be improved by 
a sophisticated shell-model calculation with 
configuration-mixing \cite{Umeya08}. 

\Figuretable{TABLE I}

The early theoretical prediction 
by Tretyakova and Lanskoy \cite{Tretyakova03} 
differs from the present result.
They have shown that the $\Sigma^-$-mixing probabilities 
on $p$-shell nuclei are the order of $10^{-3}$-$10^{-2}$ \%, 
which are smaller than ours by one or more orders of magnitude, 
within Hartree-Fock single-particle calculations
based on two-body $\Lambda N$-$\Sigma N$ effective 
interactions \cite{Akaishi00}.
The $\Lambda$-$\Sigma$ coupling in the Hartree-Fock states 
seems to be hindered by the lack of active configurations. 
For the two-step mechanism, $\pi^-p$ $\to$ $\pi^0n$ followed by 
$\pi^0p$ $\to$ $K^+\Lambda$ or  
$\pi^-p$ $\to$ $K^0\Lambda$ followed by $K^0p$ $\to$ $K^+n$, 
in the DCX $^{10}$B($\pi^-$,~$K^+$) reaction, 
we roughly estimate the integrated lab cross sections of 
$d\sigma/d\Omega$ for the $^{10}_{~\Lambda}$Li bound state 
with a harmonic oscillator model 
in the eikonal approximation \cite{Chrien92}.
In Table~\ref{tab:table1}, we show that the calculated value of 
$d\sigma/d\Omega$ at 1.20 GeV/c (6$^\circ$) 
by the two-step mechanism is rather small (1-2 nb/sr) 
compared to that by the one-step one. (See also Fig.~\ref{fig:2}.) 
The incident-momentum dependence of $d\sigma/d\Omega$
in the data is similar to that in the one-step mechanism.
Therefore, we believe that the one-step mechanism 
is dominant in the ($\pi^-$,~$K^+$) reaction at 1.20 GeV/c. 
The $^{10}$B($\pi^-$,~$K^+$) experiment at KEK \cite{Saha05} 
might be interpreted as a measurement of the $\Sigma^-$ 
admixture in the $\Lambda$ hypernucleus. 
The $\Sigma^-$ admixture gives a key for understanding 
of the EoS and neutron stars \cite{Myint00}.

In conclusion, 
the calculated spectrum of the $^{10}_{~\Lambda}{\rm Li}$ 
hypernucleus 
by the one-step mechanism via $\Sigma^-$ doorways 
fully explains the data of the DCX $^{10}$B($\pi^-$,~$K^+$) reaction 
at 1.20 GeV/c, rather than by the two-step mechanism. 
The result shows that the $\Sigma^-$ admixture probability in 
the $^{10}_{~\Lambda}$Li bound state is the order of 10$^{-1}$ \%.
The sensitivity to the potential parameters 
implies that the nuclear ($\pi^-$,~$K^+$) reactions 
with much less background experimentally 
provide a high ability for the theoretical analysis of 
precise wave functions in the neutron-rich 
$\Lambda$ hypernuclei.
The detailed analysis based on microscopic nuclear calculations 
is required for forthcoming J-PARC experiment \cite{Sakaguchi07}.
This investigation is in progress.


The authors are obliged to T. Fukuda, Y. Akaishi and H. Nemura 
for many discussions,
and to A. Sakaguchi and H. Noumi for useful comments.
They are pleased to acknowledge D.J. Millener for 
valuable discussions and comments. 
This work was supported by Grant-in-Aid for Scientific Research on
Priority Areas (No. 17070002 and No. 20028010). 



\clearpage

\clearpage

\begin{table}[tb]
\caption{
\label{tab:table1}
Calculated results of the integrated lab cross sections of
$d\sigma/d\Omega$ for the $^{10}_{~\Lambda}$Li 2$^-$ bound state 
with two-step and one-step processes in $^{10}$B($\pi^-$,~$K^+$) 
reactions at 6$^\circ$, 
compared with the data \cite{Saha05}. 
The value in the bracket 
is a lower limit one with $\Lambda$ quasi-free corrections. 
}
\begin{ruledtabular}
\begin{tabular}{ccccc}
$p_{\pi}$   & Two-step\footnotemark[1] 
            & One-step\footnotemark[2] 
            & Exp. \cite{Saha05}  \\
  (GeV/c)  &  (nb/sr) &  (nb/sr) &  (nb/sr)  \\
\hline
  1.05  &  $\sim$1.6 & 2.4  & 5.8$\pm$2.2\footnotemark[3]   \\
  1.20  &  $\sim$1.2 & 5.4  & 11.3$\pm$1.9\footnotemark[3] (9.6$\pm$2.0) \\
\end{tabular}
\end{ruledtabular}
\footnotetext[1]{Sum of the cross sections via
$\pi^-p$ $\to$ $\pi^0n$ followed by $\pi^0p$ $\to$ $K^+\Lambda$ 
and  
$\pi^-p$ $\to$ $K^0\Lambda$ followed by $K^0p$ $\to$ $K^+n$, 
by a simple harmonic oscillator model.}
\footnotetext[2]{
$P_{\Sigma^-}=0.57$ \% ($V_{\Sigma\Lambda}=$ 11 MeV) is assumed.}
\footnotetext[3]{All the events for $-20$ MeV $\le E_\Lambda \le$ 0 MeV.}
\end{table}

\end{document}